# The operational and control software of Multi-channel Antarctic Solar Telescope

Ya-qi Chen, Ming-hao Jia, Guang-yu Zhang, Zhi-yue Wang, Yi-ling Xu, Yi Feng, Jie Chen, Hong-fei Zhang, Peng Jiang, Tuo Ji, Jian Wang, *Senior Member, IEEE*

*Abstract*—MARST (Multi-channel Antarctic Solar Telescope) aims to observe the Sun in multiple wavebands in Antarctica and will be China's first solar telescope in Antarctica. The telescope has two tubes, corresponding to Photosphere observation which uses 11 filters and Chromosphere observation in Hα waveband. The two tubes need to be coordinated to observe at the same time. The telescope will observe the Sun for a long time, so a self-guiding module is needed to improve sun tracking. Besides, performing solar specific flat-field exposure is necessary for analyzing. EPICS is introduced to control each hardware and an autonomous observation system based on RTS2 is designed under such demands. EPICS application modules are implemented for each device: telescope mount & focuser, filter wheel, Hα filter, dome with webcams, Andor CCD and PI CCD. We also integrate EPICS modules into RTS2 framework with an XML format configuration. To control these applications autonomously we have developed a RTS2 executor module where two plan classes are instantiated to control two sets of filters and CCDs, and to ensure only one could control the mount at the same time. Different types of observation plans are designed to describe different series of processes with different priorities. To improve sun tracking, we calculate the centroid of each image to get the offset, then apply the correction to the telescope during observation process. For frontend users, a GUI based on PyQt5 and QML is implemented and connects to rts2-httpd and rts2-proxy modules so that users can control devices, check images and get logs.

*Keywords*：*MARST, autonomous observation, RTS2, EPICS*

## I. INTRODUCTION

MARST (Multi-channel Antarctic Solar Telescope) will be China's first solar telescope in Antarctica that aims to observe solar UV and white light spectrum radiation, to study Solar Flare, Sunspot and other continuous spectrum changes (e.g. Ballmer). Influence on the earth's climate and Solar structure could also be learned. MARST is a small equatorial telescope with two tubes as shown in Figure 1. One of the tube is for Photosphere observation under multiband to study continuous spectrum which has a filter wheel, focuser and Andor CCD to control. The other tube is for Chromosphere observation in Hα bandwidth to study solar activity which has an Hα filter, focuser and Princeton Instruments CCD to control. In addition, a dome with two webcams also need to be controlled.

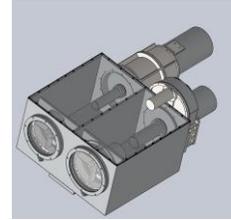

Figure 1 Diagram of Two-tube Structure

## II. SYSTEM DESIGN

The two tubes need to be coordinated to observe at the same time. The telescope will monitor the Sun for a long time, so self-guiding software module has been considered to improve mount tracking. Furthermore, a Solar specific flat-field exposure process has also been considered. Under such demands, EPICS (Experimental Physics and Industrial Control System) is chosen so that we can benefit from its soft real-time advantage to control hardware. RTS2 (Remote Telescope System, Version 2) is also adopted which is suitable to perform autonomous and remote observation. As Figure 2 shows, we design our software control system into three layers, of which from bottom to top are: device control layer, observation operation layer and user interface layer. The whole control system is developed under Linux.

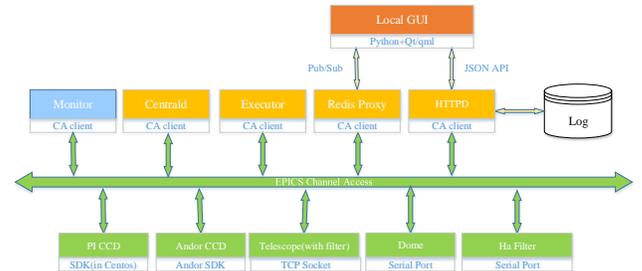

Figure 2 Diagram of Software Structure

This work was supported by the National Natural Science Funds of China under Grant No: 11603023, 11773026, 11728509, the Fundamental Research Funds for the Central Universities (WK2360000003, WK2030040064), the Natural Science Funds of Anhui Province under Grant No: 1508085MA07, the Research Funds of the State Key Laboratory of Particle Detection and Electronics, the CAS Center for Excellence in Particle Physics, the Research Funds of Key Laboratory of Astronomical Optics & Technology, CAS.

The Authors Ya-qi Chen, Ming-hao Jia, Guang-yu Zhang, Zhi-yue Wang, Yi-ling Xu, Yi Feng, Jie Chen, Hong-fei Zhang, Jian Wang is with the University of Science and Technology of China, Jian Wang, State Key Laboratory of Technologies of Particle Detection and Electronics, University of Science and Technology of China, Hefei, Anhui 230026, China (e-mail: Jian Wang, wangjian@ustc.edu.cn). The Authors Peng Jiang, Tuo Ji is with Polar Research Institute of China, Shanghai 200136, China

At device control layer, each device is controlled independently by a corresponding EPICS IOC (Input Output Controller). CCDs are controlled via USB communication, based on AreaDetector. On account of the fact that the PI CCD driver needs an old version of Linux kernel, we come up with a workaround solution by virtualizing CentOS 6 in a machine emulator called QEMU. The telescope mount, filter wheel and focusers are controlled separately as a TCP server. We use a TCP client to communicate with it following the specified agreements. Dome and Hα filter are controlled via serial port.

At observation operation layer, devices ought to be coordinated to perform complete control process. XML format configuration is introduced to define properties of devices and is parsed by RTS2. With the help of the predefined XML configuration, RTS2 modules then can find devices and map device properties to RTS2 value automatically based on EPICS Channel Access. To perform autonomous observation, a RTS2 executor module is designed and implemented. The executor monitor state changes of devices to decide the next operation. The executor mainly consists of a parser to parse command sequences, a command sequence object with tree-like command objects (serial or parallel sequence), a timer to handle observation timeout and resending, a method called 'stateChanged' to handle command execution state and RTS2 DevClients to interact with each device. An example operation is shown in Figure 3.

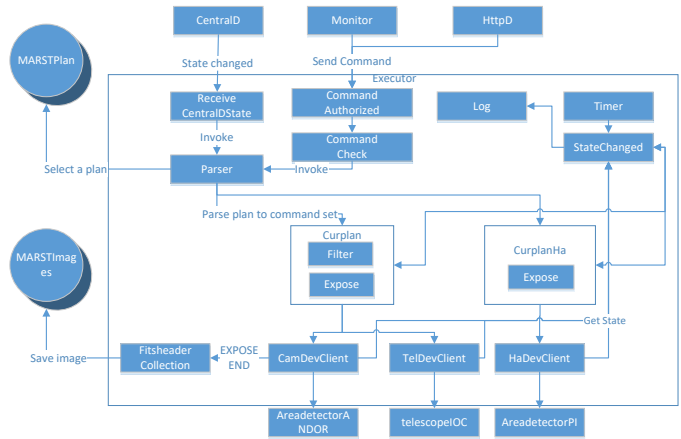

Figure 3 Diagram of Executor operation process

The observation modes of MARST include normal long-time observation, dark, flat and bias. For long-time observation mode, the executor will setup two tubes separately and wait until the mount is ready. For flat mode, the default exposure start time is set to 12:00 a.m. each day and can be modified to meet observation needs. Twelve positions of the sky are predefined to be exposed so that the Sun in the images can cover the whole field of view. Different observation modes have different priorities so that there will be only one plan executing at a time and the plan with a higher priority will be executed when conflict occurs.

In addition, for the sake of the mount's lack of tracking accuracy, a self-guiding module is developed to fix tracking error accumulation. Centroid of the image will be calculated after each exposure based on luminosity. Thus, offset between neighboring images can be calculated. The executor will suspend observation and resume the telescope if the offset is larger than limit which is considered as an exceptional event.

User interface is based on PyQt5 and QML which is easy to extend. The UI connects with rts2-httpd and Channel Access to fetch device information and can display images, send commands and monitor status as shown in Figure 4.

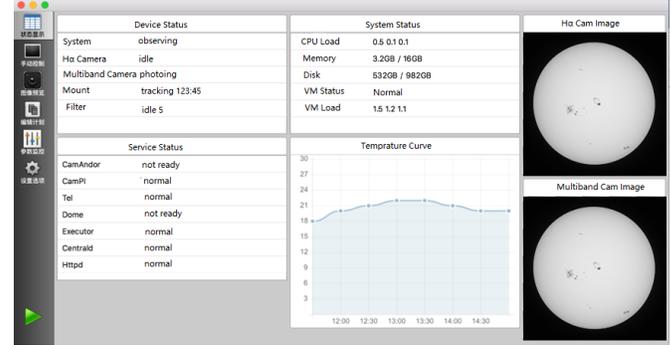

Figure 4 Diagram of User Interface

### III. SYSTEM TEST

At present the software is under test with real hardware in Nanjing, Jiangsu as shown in Figure 5. All the EPICS IOCs can function properly after large amounts of tests so that the observation processes have become the main focus right now.

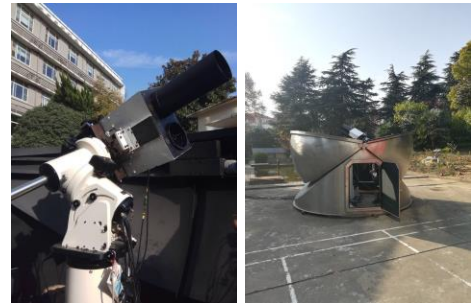

Figure 5 Test spot in Nanjing, Jiangsu, China

### IV. CONCLUSION

We have successfully implemented a software control system for MARST based on EPICS and RTS2 which is scalable enough to execute different observation processes with the help of XML format configuration.

The software control system has been successfully deployed at Nanjing Institute of Astronomical Optics and Technology, National Astronomical Observatories, CAS for daily tests and has shown its robustness after long-term operation.

REFERENCE

[1]. Guang-yu Zhang, Jian WANG, et al. "An Autonomous Observation and Control System Based on EPICS and RTS2 for Antarctic Telescopes," Monthly Notices of the Royal Astronomical Society 455 (2), 1654-1664 (2016).
[2]. Jia M., Chen Y., Zhang G., Jiang P., Zhang H., Wang J., A web service framework for astronomical remote observation in Antarctica using satellite link. Astronomy and Computing (2018), https://doi.org/10.1016/j.ascom.2018.04.005
[3]. Petr Kubánek, "RTS2—The Remote Telescope System," Advances in Astronomy, vol. 2010, Article ID 902484, 1-9 (2010)